\documentclass[submission,copyright,creativecommons]{eptcs}
\providecommand{\event}{F-IDE 2021} 
\usepackage{underscore}           

\usepackage{graphicx}

\usepackage{amsmath}
\usepackage{algorithm}
\usepackage[utf8]{inputenc}
\usepackage{amssymb}
\usepackage[noend]{algpseudocode}
\usepackage[english]{babel}	
\usepackage{paralist}	
\usepackage{listings}
\usepackage{color}
\usepackage{hyperref}
\usepackage{lineno}
\usepackage{amssymb}
\usepackage{dafny}
\usepackage[colorinlistoftodos]{todonotes}
\usepackage{ulem}

\title{Verifying Time Complexity of Binary Search using Dafny}

\author{Shiri Morshtein
\institute{School of Computer Science \\The Academic College Tel-Aviv Yaffo}
\email{shirim66@gmail.com}
\and 
Ran Ettinger
\institute{Department of Computer Science \\ Ben-Gurion University of the Negev}
\email{ranger@cs.bgu.ac.il}
\and 
Shmuel Tyszberowicz
\institute{RISE - Centre for  Research and Innovation in Software Engineering \\Southwest University, Chongqing, China}
\institute{Department of Software Engineering, Afeka Academic College of Engineering\\ Tel Aviv, Israel}
\email{tyshbe@tau.ac.il}
}
\def\titlerunning{Verifying Time Complexity}
\def\authorrunning{Morshtein, Ettinger and Tyszberowicz}

\begin{document}
%
%
%


%
%
\maketitle              
\begin{abstract}
Formal software verification techniques are widely used to specify and prove the functional correctness of programs. However, nonfunctional properties such as time complexity are usually carried out with pen and paper. Inefficient code in terms of time complexity may cause massive performance problems in large-scale complex systems. We present a proof of concept for using the Dafny verification tool to specify and verify the worst-case time complexity of binary search. This approach can also be used for academic purposes as a new way to teach algorithms and complexity.

\end{abstract}

\section{Introduction}\label{1}
\begin{figure}
\centering
\includegraphics[scale=1]{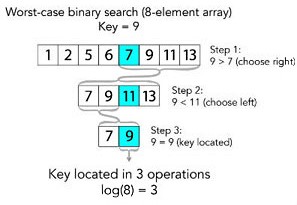}
\caption{A running example of binary search.} \label{fig1}
\end{figure}

The binary search algorithm is a well-known example used in courses such as Algorithms and Data Structures to demonstrate the logarithmic time complexity search algorithm, searching a value in a sorted array of elements. We use a sorted sequence as that data structure where we search the key.
In each iteration, the middle element of an interval is tested, and if it is not the required key, half of the sequence in which the key cannot lie is eliminated, and the search continues on the remaining half. It is repeated until either the key is found or the remaining half is empty, which means that it is not in the sequence. 
Since each iteration narrows the search range by half, after k steps, the algorithm reduces the range by $2^k$. As soon as $2^k$ equals or exceeds n, the process terminates. Given that $n \leq 2^k$ is equivalent to $log_2n \leq k$, the process terminates after at most $log_2n$ steps.
Figure~\ref{fig1} presents a running example of the algorithm.

It is not trivial to establish the correctness of the algorithm, because the index calculations might lead to programming errors~\cite{[48],[2]}. In our implementation of the algorithm, $q$ is an input sequence, $|q|$ denotes its size, where the indices range from 0 to $|q|-1$. 
Initially, we assign the return value $r$ the value -1. 
If the key is found, we change the value to the relevant index, which must range between 0 to $|q|-1$; otherwise, -1 is returned. Hence, to specify the functional requirements of the algorithm, the following expressions must hold:
\begin{equation} \label{eq1}
	0 \leq r \Rightarrow (r <  |q| \text{ and } q[r]=key) 
\end{equation}
\begin{equation} \label{eq2}
	r < 0 \Rightarrow \text{ key !in q }
\end{equation}

In addition, to specify the requirement that the binary search time complexity belongs to $O(log_2n)$, we define $T(n)$---the binary search time complexity, where T is the time complexity of binary search on input of size $n=|q|$. 
\begin{equation} \label{eq3}
	T(n)=O(log_2~n) 
\end{equation}
i.e.,\
\begin{equation} \label{eq4}
	\exists c>0~and~\exists n_0>0~s.t.~\forall n. \text{ }n \geq n_0 \Rightarrow \text{$T(n)$} \leq c\cdot log_2n
\end{equation}
In this paper we demonstrate the use of Dafny~\cite{[5]} for specifying and proving time
complexity of the binary search algorithm; i.e.,\ our main goal is to prove that the binary search time complexity belongs to $O(log_2n)$. Dafny is an imperative language that supports formal specification using annotations. It builds on the Boogie intermediate language~\cite{[4]}, which uses the Z3 automated theorem prover for discharging proof obligations~\cite{[3]}. By defining theorems and using pre- and postconditions (\textit{requires} and \textit{ensures}, respectively), loop invariants (\textit{invariant}), assertions (\textit{assert}), and other constructs, we build fully verified programs. Cases where the theorem cannot be proved automatically, may require some help from the developer. The developer may assist Dafny, usually by providing \textit{assertions} or \textit{lemmas}~\cite{[6]}. 
Dafny is an auto-active program verifier~\cite{[42]}, which is more automated than proof assistants such as Coq~\cite{[34]}. It provides the platform to write a verified version of an algorithm, including the implementation and verification in the same method. 

Section~\ref{2} provides the definition, declaration, and specification of the \textit{big-O} notation for logarithmic functions, implementing it by creating mathematical definitions for O($log_2n$), followed by the verification of the functional correctness of the binary search algorithm. 
The challenges of using Dafny to create the nonfunctional specifications and proofs are presented in Section~\ref{3}, that also illustrates a process and a methodology of proving an approximated time boundary for an algorithm, proving that the respective time boundary belongs to O($log_2n$).
This process presents an approach for computer science students and software developers to design and implement verified programs on a functional measure and for time complexity assessment. Our main contribution is the addition of the time complexity big-O definition to the specification and verification process. The methodology of defining and proving the correctness of the time complexity with Dafny (presented in the sections \ref{4}, \ref{5}, \ref{6}), arms the user with deeper understanding of time complexity big-O notation of the algorithm and assists with the implementation of efficient programs. The advantages of this approach compared to previous work are discussed in Section~\ref{7}. The paper concludes in Section~\ref{8}.

The code that is presented in this paper is available at:\\
\url{https://github.com/shirimo/binary-search-verification}~ and was run with Dafny 3.1.0: \url{https://github.com/dafny-lang/dafny/releases/tag/v3.1.0}~.

\section{Specification of Logarithmic Complexity Class}\label{2}
The big-O notation is a formal way to express the upper bound of an algorithm's runtime. 
Since functions in Dafny define mathematical functions, \textit{predicates} (Boolean functions) are used to define conditions that those functions must fulfill.
The function $f$, belongs to O($log_2n$) if:
\begin{equation}\label{eq6}
	\exists c>0~and~\exists n_0>0~s.t.~\forall n. \text{ }n \geq n_0 \Rightarrow \text{$f(n)$} \leq c\cdot log_2n
\end{equation}

In this case, $log_2n$ is a function that must be defined for Dafny.
\label{lst:Log2}
\begin{lstlisting}[language=Dafny, basicstyle=\small, tabsize=2]
function Log2 (n: nat): nat
	requires n>0
	decreases n
{
	if n==1 then 0 else 1 + Log2(n/2)
}
\end{lstlisting}
The \textit{nat} type stands for the natural numbers (non-negative integers), which is required by the $log_2n$ function. Also, the $log_2n$ function must require a positive input, hence the $n>0$ precondition.
The \textit{decreases} annotation is used to assist Dafny to prove that the function terminates and calls itself with a valid value. It provides an expression that decreases with every recursive call and is bounded by 0.
\textit{Log2} holds, since each recursive call divides the value by two, starting with a non-negative value. Thus, it necessarily becomes smaller with each call.

Now that the $log_2n$ function is defined, we specify the O($log_2n$). 

The predicates and the lemma that define O($log_2n$) are:
\begin{lstlisting}[language=Dafny, basicstyle=\small, tabsize=2]
predicate IsOLog2n(n: nat ,t: nat)
{
	exists f: nat -> nat :: t<=f(n) && IsLog2(f)
}
predicate IsLog2 (f: nat -> nat)  
{
	exists c :nat, n0: nat :: IsLog2From(c, n0, f)
}
predicate IsLog2From (c :nat, n0: nat, f: nat -> nat)
{
	forall n: nat {:induction}:: 0 < n0 <= n ==> f(n) <= Log2(n)*c
}
lemma logarithmic (c :nat, n0: nat, f: nat -> nat)
	requires IsLog2From(c, n0, f)
	ensures IsLog2(f)
{}
\end{lstlisting}

The predicate \textit{IsOLog2n} is required to prove that such a function \textit{f} which satisfies expression~(\ref{eq3}) and upper-bounds the algorithm's runtime indeed exists, where n is the size of the input.
It defines the postcondition we strive to establish for any logarithmic algorithm.
The predicates and the lemma define the mathematical expression of O($log_2n$) as defined in expression~(\ref{eq6}):
The \textit{IsLog2From} predicate determines if for given positive $c,n_0$, for every n larger than n0, $f(n)$ is upper-bounded by $c*log_2n$ (the second part of expression~(\ref{eq6})).
The \textit{IsLog2} predicate determines if such positive $c,n_0$ that satisfies \textit{IsLog2From} exist (the first part of expression~(\ref{eq6})). 
The \textit{logarithmic} lemma is provided to merge the two predicates above it to the complete definition.
The developer does not need to provide further proof to deduce that if the inputs \textit{f,c,n0} satisfy the predicate \textit{IsLog2From}, then \textit{f} satisfies the predicate \textit{IsLog2} and hence \textit{f} is logarithmic.

\section{Functional Specification and Verification}\label{3}
We now show a fully verified implementation of the binary search algorithm.
The use of binary search  requires that the input sequence is sorted:
\begin{lstlisting}[language=Dafny, basicstyle=\small, tabsize=2]
predicate Sorted (q: seq<int>)
{
	forall i, j :: 0<=i<=j<|q| ==> q[i]<=q[j]
}
\end{lstlisting}

For verifying the functional requirement of binary search, as defined in Section~\ref{1}, we need to ensure both expressions~(\ref{eq1}) and~(\ref{eq2}) are satisfied:
\begin{lstlisting}[language=Dafny, basicstyle=\small, tabsize=2]
predicate BinaryPosts (q: seq<int>, r: int, key: int)
{
	(0<=r ==> (r<|q| && q[r]==key)) && 
	    (r<0 ==> key !in q)
}
\end{lstlisting}
Now the binary search method can be declared with the functional specification: 
\begin{lstlisting}[language=Dafny, basicstyle=\small, tabsize=2]
method binarySearch (q: seq<int>, key: int) returns (r: int)
	requires Sorted(q) 
	ensures BinaryPosts(q, r, key)
\end{lstlisting}
\noindent These properties can be used to prove the correctness of the search. The method body is given below:
\begin{lstlisting}[language=Dafny, basicstyle=\small, tabsize=2]
r := -1;
var lo, hi := 0, |q|;
while lo < hi
	invariant BinaryLoop(q, lo, hi, r, key)
	decreases hi-lo
{
	var mid := (lo+hi)/2;
	if key < q[mid] { hi := mid; }
	else if q[mid] < key { lo := mid + 1; } 
	else 
	{
		r := mid;
		hi := lo;
	}
}
\end{lstlisting}
Notice, Dafny uses mathematical integers, not the bounded integers found in most popular programming languages. It means that there is no issue of overflow in Dafny \cite{[40]} in the operation
\textit{(hi+lo)/2}.

A loop invariant is an expression that holds upon entering a loop and after every loop body execution. The loop invariants of the binary search algorithm are grouped into a predicate for a cleaner implementation.
Note that in Dafny $q[..lo]$ does not include the $lo$ index.
\begin{lstlisting}[language=Dafny, basicstyle=\small, tabsize=2]
predicate BinaryLoop (q: seq<int>, lo: int, hi: int, 
    r: int,key: int)
{
	(0<=lo<=hi<=|q| && 
	(r<0 ==> key !in q[..lo] && key !in q[hi..]) && 
	(0<=r ==> r<|q| && q[r]==key)
}
\end{lstlisting}
 
This predicate is composed of three expressions that must hold for every iteration: the first 
defines that indices' variables are within the input sequence; the second is concluded from the first postcondition---if  \textit{r} is -1, the key has not been found yet, so it does not lie in the part of the sequence that we already had eliminated; the
third is the same as the second postcondition; if the returned value \textit{r} is non-negative, the key has been found, and r is assigned with its location. 
As we can see, when the loop invariant holds, the postcondition is satisfied; thus, the functional properties
of the algorithm are verified.

\section{Binary Search Logarithmic Time Complexity Specification}
\label{4}
To specify the binary search algorithm's runtime, we define the time complexity value with a \textit{ghost variable t}. It counts the number of dominant operations performed by the algorithm~\cite{[18]} (i.e.,\ the number of loop iterations), and it is added to the algorithm's postcondition. The postcondition includes the expression that must hold for logarithmic time complexity as defined and detailed in Section~\ref{2}. These \textit{ghost} entities in Dafny are used only during verification and are excluded from the executable code.
\begin{lstlisting}[language=Dafny, basicstyle=\small, tabsize=2]
method binarySearch (q: seq<int>, key: int)
  returns (r: int, ghost t: nat)
	requires Sorted(q) 
	ensures BinaryPosts(q, r, key)
	ensures |q|>0 ==> IsOLog2n(|q|,t)
	ensures |q|==0 ==> t==0
{
	t := 0;
	r := -1;
	if |q|>0
	{
		var lo, hi := 0, |q|;
		while lo < hi
			invariant BinaryLoop(q, lo, hi, r, key)
			decreases hi-lo
		{
			...
			t := t+1;
		}
	}
}
\end{lstlisting}
Here the time complexity variables are inserted into the algorithm's specification.
Dafny fails to verify the postcondition 
\begin{lstlisting}[language=Dafny, basicstyle=\small, tabsize=2]
 ensures|q|>0 ==> IsOLog2n(|q|,t), 
\end{lstlisting}
complaining that the postcondition might not hold, due to the non-satisfied expression represented by the predicate \textit{IsOLog2n}:
\begin{lstlisting}[language=Dafny, basicstyle=\small, tabsize=2]
predicate IsOLog2n (n: nat ,t: nat )
{
	exists f: nat -> nat :: t<=f(n) && IsLog2(f)
}
\end{lstlisting}
This predicate would be satisfied when we prove such logarithmic function \textit{f}, that upper-bounds the binary search algorithm's time complexity, exists. This proof can be done by the following lemma:
\begin{lstlisting}[language=Dafny, basicstyle=\small, tabsize=2]
lemma OLog2nProof (n: nat, t: nat)
	requires n>0
	requires t<=f(n)
	ensures IsOLog2n(n, t)
{
	var c, n0 := logarithmicCalcLemma(n);
	logarithmic(c, n0, f);
}
\end{lstlisting}
This lemma requires our assumed function \textit{f}, as a global variable. The \textit{logarithmicCalcLemma} is calculating the values of \textit{c,n0} that are required to prove that \textit{logarithmic(c,n0,f)}. The lemma specification is:
\begin{lstlisting}[language=Dafny, basicstyle=\small, tabsize=2]
lemma logarithmicCalcLemma(n: nat) returns(c: nat, n0: nat)
	requires n>0
	ensures IsLog2From(c, n0, f)
\end{lstlisting}
The lemma body is explicitly written for the function \textit{f} in Section~\ref{6}.
The following sections are engaged in deriving this function $f$.

\section{Derivation of a tight Upper-Bound Function}\label{5}
We now move from intuition to mathematics. To prove that $t\leq f(n)$, a suitable logarithmic function f must be provided. To define a function that upper-bounds binary search and is expressed with the \textit{Log2} function, we analyze the behavior of the binary search algorithm's loop in relation to the \textit{Log2} [\ref{lst:Log2}] function:
\begin{lstlisting}[language=Dafny, basicstyle=\small, tabsize=2]
var lo, hi := 0, |q|;
while lo < hi
	invariant BinaryLoop(q, lo, hi, r, key)
	decreases hi-lo
{
	var mid := (lo+hi)/2;
	if key < q[mid] { hi := mid; }
	else if q[mid] < key { lo := mid + 1; } 
	else {
		r := mid;
		hi := lo;
	}
	t := t+1;
}
\end{lstlisting}

It is hard for the developer, and for Dafny, to perform the transition from an iterative implementation as in \textit{binary search} to a recursive one as in \textit{Log2}. That is why we reduce the problem as follows:
\begin{enumerate}[(a)]
\item \label{a}Defining a transition recursive function which imitates the actions performed in the binary search method, returning the number of recursive calls rather than the key's index or -1.
\item \label{b}Proving, by adding loop invariants to the binary search method, that the transition function upper-bounds the value of \textit{t}.
\item \label{c}Deriving a function that upper-bounds the transition function and is expressed with \textit{Log2}.
\end{enumerate}
Following the respective steps, we transitively prove that the logarithmic function upper-bounds the binary search estimated runtime.

\bigskip
\noindent
Step (\ref{a}): \label{5(a)} \label{lst:TBS}
\begin{lstlisting}[language=Dafny, basicstyle=\small, tabsize=2] 
function TBS (q: seq<int>, lo: int, hi: int, key: int): nat
	requires 0<=lo<=hi<=|q|
	decreases hi-lo
{
	var mid := (lo+hi)/2;
	if hi-lo==0 || |q|==0 then 0 
	else if  (key==q[mid] || hi-lo==1) then 1 
	else if key<q[mid] then 1 + TBS(q, lo, mid, key) 
	else 1 + TBS(q, mid+1, hi, key)
}
\end{lstlisting}
This function imitates the actions performed by the binary search algorithm, with the same range of indices and decreases definition, implying the same number of steps as the binary search algorithm's loop. For now, we assume that \textit{TBS} returns the same value as the number of steps the binary search algorithm's loop performs. Next, the focus is on the formal verification made by Dafny to prove the correctness of this assumption.

\bigskip
\noindent
Step (\ref{b}): \label{5(b)}
\begin{lstlisting}[language=Dafny, basicstyle=\small,tabsize=2]
while lo < hi
	invariant BinaryLoop(q, lo, hi, r, key)
	invariant t <= TBS(q, 0, |q|, key) - TBS(q, lo, hi, key)
	decreases hi-lo
{
    ...
    t := t+1;
}
\end{lstlisting}
Here the time complexity variables are inserted into the loop's specification. To prove that \textit{TBS} returns the same value as the number of steps the binary search algorithm's loop performs (\textit{ghost var t}), the correctness of the assumption in step (a) must be proven in each loop iteration. Hence, the loop invariant \textit{$t \leq TBS(q,0,|q|,key) - TBS(q,lo,hi,key)$} has been added. This loop invariant holds, thus each decision made during the loop body is made also by the \textit{TBS} function. That is, for each iteration, the step counter \textit{t} must be upper-bounded by the difference between the algorithm's runtime for the whole sequence to the current part of the sequence. The loop ends when \textit{$lo=hi$}, where the \textit{TBS} function returns 0. Therefore, the method terminates when the loop invariant holds for $t \leq TBS(q,0,|q|,key)$, and that gives us an upper-bound to the algorithm's time complexity. In the next section we specify and verify several lemmas to ensure that \textit{TBS} is logarithmic.

\bigskip
\noindent
Step (\ref{c}): \label{5(c)}
\noindent
In this step we calculate the time complexity of the logarithmic function. As a pen and paper calculation would have been done, we analyze the \textit{TBS} [\ref{lst:TBS}] function in relation to the \textit{Log2} [\ref{lst:Log2}] function:

We start by analyzing the stop condition. The function \textit{TBS} stops and returns~0 when the inspected size is 0 (regardless if the size of the sequence is 0 or the difference between \textit{hi} and \textit{lo} is 0). 
However, \textit{Log2} stops and returns 0 for the input 1. Therefore, 1 is added to the mathematical expression of \textit{TBS}. 
Continuing with the analysis of the value divided by 2, \textit{TBS} gets the value \textit{lo+hi}, summing 2 sequence indices. 
This sum can at most be 2*$|q|$-1 (when hi is $|q|$ and lo is $|q|-1$). 
Therefore, the mathematical expression of \textit{TBS} must be doubled. 
The last thing 
to consider is that the input for \textit{Log2} is greater or equal 1. 
The behavior of the rest of the functions is similar. In each call only a part of the initial size is assigned and the result increases by 1. The result is the expression:
\begin{equation} \label{eq7}
	n>0 \Rightarrow \text{($TBS(q,lo,hi,key) \leq 2*log_2(|q|+1) + 1$)}
\end{equation}
This analysis has also been done the same way as pen and paper analysis. That being so, we now provide a proof that the analysis holds with the following lemma:
\begin{lstlisting}[language=Dafny, basicstyle=\small,tabsize=2]
lemma TBSisLog (q: seq<int>, lo: nat, hi: nat, key: int)
	requires |q|>0
	requires 0 <= lo < hi <= |q|
	decreases hi-lo
	ensures TBS(q, lo, hi, key) <= 2*Log2(hi-lo)+1
{
	var mid := (lo+hi)/2;
	if key<q[mid] && 1<hi-lo
	{
		TBSisLog(q, lo, mid, key);
	}
	else if key>q[mid] && hi-lo>2
	{
		log2Mono(hi-(mid+1), (hi-lo)/2);
		TBSisLog(q, mid+1, hi, key);
	}
}
\end{lstlisting}
This lemma adds the missing requirements for enabling the use of the log function: the precondition requires the sequence is non empty, and the \textit{lo} index to be strictly smaller than the \textit{hi} index. The \textit{ensures} expression is the proof obligation as in expression (\ref{eq7}). This lemma calls itself recursively. The recursive call is treated in accordance with programming rules: the precondition of the callee is checked, termination is checked, and the postcondition can be assumed. In effect, this sets up a proof by induction, where the recursive call to the lemma acts as an inductive step.
The lemma performs the same actions \textit{TBS} does, excluding cases where the new preconditions ($lo < hi$ \textit{and} $0 < |q|$) are violated.
The \textit{log2Mono} lemma is for Dafny to understand that \textit{Log2} is a monotonic function.
\begin{lstlisting}[language=Dafny, basicstyle=\small, tabsize=2]
lemma log2Mono (x: nat, y: nat)
	requires x>0 && y>0
	ensures y>=x ==> Log2(y)>=Log2(x)
	decreases x, y 
{
	if x!=1 && y!=1 {log2Mono(x-1,y-1);}
}	
\end{lstlisting}
The \textit{log2Mono} will be used again for the \textit{Log2} function's requirements. Since a natural number can be of value 0 in our context, the upper-bounded function  is increased from ($2*log_2(|q|)+1$)  to  ($2*log_2(|q|+1)+1$). 
Now it has been proven transitively that the binary search algorithm has a mathematical function, expressed with \textit{Log2}, that upper-bounds the variable \textit{t} and verified by Dafny with the help of the lemmas \textit{log2Mono} and  \textit{TBSisLog}. To make sure Dafny verifies this upper-bound for \textit{t}, we insert a temporary \textit{assertion} at the end of the method. This assertion will later be replaced by a lemma.
\begin{lstlisting}[language=Dafny, basicstyle=\small, tabsize=2]
method binarySearch (q: seq<int>, key: int) 
    returns (r: int, ghost t: nat)
	requires Sorted(q) 
	ensures BinaryPosts(q, r, key)
	ensures |q|>0 ==> IsOLog2n(|q|, t)
	ensures |q|==0 ==> t==0
{
	t := 0;
	r := -1;
	if |q|>0
	{
		var lo, hi := 0, |q|;
		while lo < hi
			invariant BinaryLoop(q, lo, hi, r, key)
			invariant t <= TBS(q, 0, |q|, key) - 
			               TBS(q, lo, hi, key)
			decreases hi-lo
		{
			var mid := (lo+hi)/2;
			if key < q[mid] { hi = mid; } 
			else if q[mid] < key {  lo := mid + 1; } 
			else 
			{
				r := mid;
				hi := lo;
			}
			t := t+1;
		}
		TBSisLog(q,0,|q|,key);
		log2Mono(|q|,|q|+1);
	}
	assert t <= 2*log2(|q|+1)+1;
}
\end{lstlisting}
Dafny still fails to verify the postcondition \textit{$|q|>0 \Rightarrow IsOLog2n(|q|,t)$}, since it has not yet been proven that $f(n)=2*log_2(n+1) + 1$ is logarithmic. 
The lemmas that have been specified in Section~\ref{4} are used to prove it.

\section{Binary Search Upper-Bound is Logarithmic}\label{6}
So far we proved that the binary search runtime upper-bound is expressed with \textit{Log2}. We now prove that this expression is $O(log_2n)$ using the lemmas \textit{OLog2nProof} and \textit{logarithmicCalcLemma}.
This requires us to find proper \textit{c} and \textit{n0} that satisfy 
expression~(\ref{eq6}). For implementing the body of \textit{logarithmicCalcLemma} we use Dafny's \textit{calc} construct~\cite{[17]}, a theorem established by a chain of formulas, each transformed into the next. That is a replica of the pen and paper complexity proof, with a correctness guarantee by Dafny:
\begin{lstlisting}[language=Dafny, basicstyle=\small, tabsize=2]
lemma logarithmicCalcLemma (n: nat) returns (c :nat, n0:nat)
	requires n>0
	ensures IsLog2From(c,n0,f)
{	
	calc {
		f(n);
	==
		2*log2(n+1) + 1;
	<=
		2*log2(n+1) + log2(n+1);
	==
		3*log2(n+1);
	<=  {assert n>=1; log2Mono(n+1,2*n);}
		3*log2(2*n);
	== 
		3*(1+log2(n)); 
	}
	assert f(n)<= 3*(1+log2(n));
	assert n>=2 ==> (f(n)<=6*log2(n));
	c, n0 := 6, 2;
}
\end{lstlisting}
This lemma works through the steps that are required to infer proper \textit{c}, \textit{n0} for $f(n) = 2*log_2(n+1) + 1$. Since this lemma ensures that \textit{f} is logarithmic, the conditions of  the \textit{OLog2nProof} lemma now hold.
\begin{lstlisting}[language=Dafny, basicstyle=\small, tabsize=2]
function f (n: nat) : nat
{
	2*Log2(n+1) + 1	
}
\end{lstlisting}
\begin{lstlisting}[language=Dafny, basicstyle=\small, tabsize=2]
lemma OLog2nProof (n: nat, t: nat )
	requires n>0
	requires t<=f(n)
	ensures IsOLog2n(n, t)
{
	var c, n0 := logarithmicCalcLemma(n);
	logarithmic(c, n0, f);
}
\end{lstlisting}
This lemma can be used now, since the implementation has already been limited to non-empty inputs and proved that the steps counter t value is at most the value of the function \textit{f} in Section~\ref{5}(c). Also, $f(n)$ is proven to be logarithmic by the definitions that have been determined on Section~\ref{2}.

Finally, the proof that the binary search algorithm's runtime is logarithmic is complete:
\begin{lstlisting}[language=Dafny, basicstyle=\small, tabsize=2]
method binarySearch (q: seq<int>, key: int) 
    returns (r: int, ghost t: nat)
	requires Sorted(q) 
	ensures BinaryPosts(q, r, key)
	ensures |q|>0 ==> IsOLog2n(|q|, t)
	ensures |q|==0 ==> t==0
{
	t := 0;
	r := -1;
	if |q|>0
	{
		var lo, hi := 0, |q|;
		while lo < hi
			invariant BinaryLoop(q, lo, hi, r, key)
			invariant t <= TBS(q, 0, |q|, key) - 
			               TBS(q, lo, hi, key)
			decreases hi-lo
		{
			var mid := (lo+hi)/2;
			if key < q[mid] { hi := mid; } 
			else if q[mid] < key { lo := mid + 1; } 
			else {
				r := mid;
				hi := lo;
			}
			t := t+1;
		}
		TBSisLog(q,0,|q|,key);
		log2Mono(|q|,|q|+1);
		OLog2nProof(|q|,t);
	}
}
\end{lstlisting}
The postcondition \textit{IsOLog2n} sums up all the required terms for having $O(log(n)$ time complexity as defined in expression \ref{eq3}. Now our process is complete.

\section{Related Work}\label{7}

Formalization of teaching mathematics through formal verification by computers is introduced and researched in~\cite{[15]},\cite{[21]}. The presented work in~\cite{[15]} experiments on formalizing mathematical proofs using the Lean theorem prover~\cite{[16]}. The experiment concludes that mathematicians with no computer science training can become proficient users of a proof assistant using formal verification. 
An approach for teaching the material of a basic Logic course to undergraduate Computer Science students, tailored to the unique intuitions and strengths of this cohort of students, is presented in~\cite{[21]}.
Our approach supports this objective and lifts it to more practical levels; by teaching the mathematics behind a complexity class and the methodology of proving that an algorithm belongs to the complexity class.

The use of formal methods in software development education is described, e.g.,\  in~\cite{[7]} and~\cite{[8]}. Both approaches are dedicated to verifying functional correctness while we can illustrate the time complexity of algorithms to developers.

A use of a powerful tool for termination analysis and extending its approach for complexity analysis is demonstrated in~\cite{[19]}. This approach is limited to Java programs, has a steep learning curve, and involves more off-the-shelf solvers.

A verification of binary search that includes time complexity verification \textit{VeriFun} \cite{[14]} has been demonstrated in~\cite{[2]}. Since VeriFun is a semi-automated system for verifying statements about programs written in a functional programming language, the loop's binary search implementation was written recursively. 
We use Dafny, which supports procedural and functional implementation.

The functional and time complexity analysis approach is presented in~\cite{[9]}, which is focused on creating a designated functional language. This approach applies only to functional programming implementation of algorithms. Using Dafny, we can generate the desired algorithm for several languages.

The motivation and concepts of the work of Gu\'eneau et al. (\cite{[20]},\cite{[10]},\cite{[36]}) enable robust and modular proofs of asymptotic complexity bounds along with their functional correctness. The methodology is built on specifying the intended behavior of a program, proving a theorem relating concrete code to the specification, and using the proof assistant \textit{Coq}~\cite{[12]}, \cite{[15]} to mechanically check every step of the proof. This methodology is based on the Coq proof assistant; it is highly expressive, requires more expertise in verification and less automated~\cite{[17]}. Coq can automatically extract executable programs from specifications to OCaml, which is functional and imperative, but the Coq specifications must be purely functional. Our methodology is based on Dafny, which supports functional and imperative programs~\cite{[37]}. Furthermore, Dafny's program verifier is more automated than Coq~\cite{[34]} and provides the platform to write a verified version of an algorithm, including the implementation, the functional verification, and the time complexity in the same method. The target audience of our methodology is computer science students and algorithms developers. This audience is not necessarily an expert in verification, so we must provide a tool that can be easily assimilated.
The approach and methods presented in~\cite{[11]} are based on the Coq proof assistant as well. The proof obligations establish the correctness of the functions and establish a basic running time result, but not an asymptotic one in terms of big-$\mathcal{O}$.  

In \cite{[35]} a framework for verifying asymptotic time complexity of imperative programs is presented. It is done by extending Imperative HOL and its separation logic with time credits. The authors have stated that part of their future work goals is to include reasoning about while and for loops (both functional correctness and time complexity) and build a single framework in which all deterministic algorithms typically taught in the undergraduate study can be verified straightforwardly. Our work fulfills both of these goals.

The methods of work in \cite{[20]},\cite{[10]},\cite{[36]}, \cite{[11]}, \cite{[35]} presented above, implemented with proof assistants. As we mentioned earlier in Section~\ref{1}, Dafny, as an auto-active program verifier~\cite{[42]} provides automation by default, and the interaction with such a verifier happens at the level of the programming language, which has the advantage of being familiar to the programmer.

\section{Conclusion}\label{8}

We investigated the addition of the nonfunctional complexity property to the specification of algorithms. Using Dafny, this property can then be verified alongside the functional properties. The concept was demonstrated on an algorithm of a non-trivial time-complexity class, namely
a logarithmic one. 

Reflecting on the presented process, it appears that a methodology emerges for the specification and verification of time-complexity properties of algorithms. The methodology involves the following steps:
\begin{enumerate}
    \item\label{Step 1} \textit{Defining big-O notation.} 
    This is a function of both runtime and input size of the algorithm.
    \item\label{Step 2} \textit{Verifying the correctness of the algorithm.} 
    To simplify the calculation of the number of steps the algorithm performed, we implemented it with a single return point;
    It would be interesting to support algorithms with more than one return point.
    \item\label{Step 3} \textit{Adding specifications for time complexity verification.} 
    This is done using a ghost variable that counts the number of elementary operations.
    \item\label{Step 4} \textit{Deriving a mathematical function that upper-bounds the time complexity.} 
    \item\label{Step 5} \textit{Proving the function belongs to logarithmic complexity class.} 
    \item\label{Step 6} \textit{Integrating the time complexity elements with the code.} Adding the annotations provided by the products of Steps 4 and 5, to make the postcondition of Step 3 hold.
\end{enumerate}

Our initial investigation presented a methodology that can be used for theoretical and practical measures. It refers to the complexity theory by using this process of specifying, implementing, and verifying algorithm complexity for academical causes. 
It is also practical since it generates fully verified code, and the algorithm's declaration includes the time-complexity specification as part of designing and writing the code processes. 

This work is part of wider research on developing methods for specifying and proving time complexity of algorithms~\cite{[47]}. For example, given that \textit{method A} calls \textit{method B}, we explain how to find the time complexity of \textit{A} using the calculated upper-bound of the ghost variable \textit{t} in the postcondition of \textit{B}.

\bibliographystyle{splncs04}
\bibliography{bibly}

\end{document}